\documentclass[preprint]{aastex} 
\begin{document}

\title{The CMB Anisotropy Power Spectrum from the {\em Background Emission Anisotropy Scanning Telescope} Experiment}

\author{Ian J. O'Dwyer\altaffilmark{1},
Marco Bersanelli\altaffilmark{2},
Jeffrey Childers\altaffilmark{3,4},
Newton Figueiredo\altaffilmark{5},  
Doron Halevi\altaffilmark{3,4},
Gregory G. Huey\altaffilmark{1,6},
Philip M. Lubin\altaffilmark{3,4,7},
Davide Maino\altaffilmark{2},
Nazzareno Mandolesi\altaffilmark{8},
Joshua Marvil\altaffilmark{3,4},
Peter R. Meinhold \altaffilmark{3,4,7},
Jorge Mej\'{\i}a\altaffilmark{9}, 
Paolo  Natoli\altaffilmark{10},
Hugh O'Neill\altaffilmark{3,4},
Agenor Pina\altaffilmark{5},
Michael D. Seiffert\altaffilmark{11}, 
Nathan C. Stebor\altaffilmark{3,4,7},
Camilo Tello\altaffilmark{9},
Thyrso Villela\altaffilmark{9}, 
Benjamin D. Wandelt\altaffilmark{1,6},
Brian Williams\altaffilmark{3,7},
Carlos Alexandre Wuensche\altaffilmark{9}
}

\altaffiltext{1}{Astronomy Department, University of Illinois at Urbana-Champaign, IL 61801-3074}
\altaffiltext{2}{Physics Department, University of Milano, via Celoria 16, 20133 Milano, Italy}
\altaffiltext{3}{Physics Department, University of California, Santa Barbara, CA 93106}
\altaffiltext{4}{UC Santa Barbara Center for High Altitude Astrophyics at White Mountain}
\altaffiltext{5}{Universidade Federal de Itajub\'a, Departamento de F\'{\i}sica e Qu\'{\i}mica, Caixa Postal 50 37500-903, Itajub\'a, MG Brazil}
\altaffiltext{6}{Department of Physics, University of Illinois at
  Urbana-Champaign, IL 61801-3080}
\altaffiltext{7}{University of California, White Mountain Research Station, CA 93514}
\altaffiltext{8}{IASF-CNR sezione di Bologna, via P.Gobetti, 101, 40129 Bologna, Italy}
\altaffiltext{9}{Instituto Nacional de Pesquisas Espaciais, Divis\~ao de Astrof\'{\i}sica, Caixa Postal 515, 12245-970 -  S\~ao Jos\'e dos Campos, SP Brazil}
\altaffiltext{10}{Dipartimento di Fisica e sezione INFN, Universit\`a di Roma "Tor Vergata", Rome, Italy}
\altaffiltext{11}{Jet Propulsion Laboratory, California Institute of Technology, Oak Grove Drive, Pasadena, CA 91109}

\begin{abstract}
The Background Emission Anisotropy Scanning Telescope (BEAST) is a
2.2m off-axis telescope with an 8 element mixed Q (38-45GHz) and Ka
(26-36GHz) band focal plane, designed for balloon borne and ground based studies of the Cosmic Microwave Background.  
Here we present the Cosmic Microwave Background (CMB) angular
power spectrum calculated from 682 hours of data observed with the
BEAST instrument.  We use a binned
pseudo-$C_{\ell}$ estimator (the MASTER method). We find results that are
consistent with other determinations of the CMB anisotropy for angular
wavenumbers $\ell$ between 100 and 600.  We also perform cosmological parameter
 estimation.  The BEAST
data alone produces a good constraint on $\Omega_k\equiv 1-\Omega_{tot}=-0.074 \pm 0.070$, consistent with a flat Universe.  A joint parameter estimation analysis 
with a number of previous CMB experiments produces results consistent with previous determinations.

\end{abstract}


\section{Introduction}
Understanding the mechanisms of structure formation in the early
universe ($10<z<1000$) is one of the most important and active areas
in Cosmology today and measurements of the Cosmic Microwave Background
(CMB) anisotropy play a pivotal role in this field. 
In the framework of the standard cosmological model, the CMB radiation is 
interpreted as the blackbody radiation associated with a
hot dense phase of the Universe, when matter and radiation were in
thermal equilibrium \citep{peebles}.  On large angular scales the
CMB radiation traces the primordial power spectrum set by physical processes
during the first instants after the Big Bang. On smaller angular
scales, CMB anisotropies are influenced by factors that control the
expansion rate of the Universe and formation of large-scale structure,
such as the cosmological constant, the matter density and the
existence and nature of dark matter \citep{KT94}. By
measuring the angular power spectrum of CMB fluctuations, one can
discriminate among various competing theories that predict the
primordial mass distribution (e.g., inflation, cosmic strings and
textures, primordial isocurvature baryonic perturbations) and
understand the gravitational collapse that ultimately brought about
the formation of galaxies. Since the fluctuation amplitudes at angular
scales of a few degrees and smaller are also sensitive to the free
electron distribution, CMB measurements can also be used to determine
the ionization history of the universe.

After the release of the WMAP full-sky data \citep{wmap}, sub-orbital CMB 
anisotropy experiments are
still of high scientific interest as they can improve angular resolution
and sensitivity over limited sky regions. The Background Emission
Anisotropy Scanning Telescope (BEAST) is the only project currently
on-going which is probing a frequency range overlapping with that of WMAP, with
improved angular resolution (up to 0.38 degrees at $\sim 40$ GHz) and potentially better
sensitivity over approximately 5\% of the sky. The experiment is installed
in a conventionally accessible, high altitude site and it has so far
accomplished three observing campaigns, on which this paper is based.  
In this paper we discuss the constraints BEAST 
places on the power spectrum of CMB
anisotropies and its consistency with data taken from a subset of previous experiments (MAXIMA1 \citep{maxima}, TOCO \citep{toco98}, BOOMERANG02 \citep{boomerang02}, DASI01 \citep{dasi}, VSA1 \citep{vsa}, ACBAR1 \citep{acbar}, CBI \citep{cbi}, WMAP \citep{wmap}).

We present a brief overview of the
experiment in \S 2 and an overview of the estimator in \S 3.  \S
4 details our implementation of the estimator for the BEAST data and
\S 5 presents the power spectrum and the parameter estimation.  We
summarize the results in \S 6.

\section{The BEAST Experiment}

BEAST is a 2.2 meter off axis telescope, currently configured with an
8 element mixed Q (38-45 GHz) and Ka (26-36 GHz) focal plane, and a 
modulating flat mirror.  BEAST was designed as a high altitude balloon 
system and had two flights:  May 20-21, 2000 and October 16, 2000.  
Subsequent to the second flight BEAST was reconfigured to take
advantage of the UC White Mountain Research Station, Barcroft Station at 
an altitude of 3.8 km in the Eastern Sierra of California.  
The instrument was fully installed and 
operational at Barcroft in July, 2001, and took data nearly
continuously until December 2001 (except for weather and several
equipment failures due to power surges and lightning).  Two more weeks 
of data were obtained in February 2002.  A second data taking campaign 
proceeded in August and September of 2002.  The data used for
determining the power spectrum presented in this paper are taken from 
all three of these campaigns.

The data presented in this paper were gathered using the BEAST
telescope in a fixed elevation mode.  The telescope is kept at a fixed
 elevation near 90 degrees and the rotation of the Earth provides the
 map scanning.  This strategy results in a sky coverage which forms an annulus 
centered on the NCP.  The annulus is 9 degrees wide and is located between 
33 and 42 degrees in declination.  

Other aspects of the BEAST experiment are described in the following papers:  
The instrument is described in \citet{childers} and a more detailed discussion of
the optics can be found in \citet{optics}.  The map-making procedure is
described in \citet{map_paper} and constraints on galactic foregrounds
in \citet{foreground}.

\section{The MASTER Method}

We extract the CMB power spectrum  from the BEAST data using the
MASTER method, a binned 
pseudo-$C_\ell$ estimator \citep{whg,Hivon1}. We chose this estimator for its ease of implementation and
the flexibility it offers, which allows testing the analysis  with  
a number of cuts and filtering schemes designed to remove galactic, 
terrestrial and instrumental foregrounds.

The MASTER method is a de-biasing
scheme calibrated against Monte 
Carlo simulations. Pseudo-$C_\ell$ are calculated on the noisy maps
over the observed region on the sky with no corrections made for the
effect of this cut in terms of the  
couplings introduced between spherical harmonic modes.  The
expectation values of these Pseudo-$C_\ell$ are modeled in terms of an
ansatz which involves, as parameters, an instrumental transfer function
$F_\ell$ and a noise bias
term $N_\ell$. These terms are  estimated from Monte Carlo
simulations of CMB signal and of experimental noise.  

The signal and noise are simulated by taking separate random realizations of
pure CMB signal and realistic simulations of experimental
noise and subjecting them separately to exactly the same data processing (such
as beam smoothing, scanning, cuts in the time-ordered data, filtering,
template removal and map-making) as the real data.

The power spectra of the resulting signal and noise maps are averaged 
over the Monte 
Carlo runs to produce expectation values of the signal-only and
noise-only power spectra. These are used to compute the transfer
function and noise bias terms in the pseudo-$C_\ell$ estimator.

To the extent to which the MASTER ansatz models the expectation values
of the pseudo-$C_\ell$ and to which our Monte Carlo procedure mimics
the acquisition of the real data, we are guaranteed an unbiased power 
spectrum result.

The experimental data is now passed through the data processing pipeline and
the pseudo-$C_\ell$ are calculated.  Since the experiment covers  
only a fraction of the sky, a coupling is introduced when performing 
the spherical harmonic transforms to calculate the power spectra.  
By  calculating the mode-mode coupling kernel for the observed unmasked 
region on the sky,  it is possible to correct for this effect.  

Lastly, a binning scheme is chosen in $\ell$ for the final power 
spectrum and a number of Monte Carlo simulations containing both signal and noise
are performed.  The covariance matrix of the estimates is calculated
by computing the pseudo-$C_\ell$ estimator on these
simulations. The diagonal elements of the binned covariance matrix
are  the variances of the binned power spectrum. 

\section{Implementation of MASTER for BEAST}

In order to produce an accurate CMB power spectrum from the BEAST
data, a detailed knowledge of the experimental beam shape and pointing
is required.  A residual $\chi^2$ fit of a smoothed delta function to
maps of Cygnus A and a best fit to the flux from Cygnus A lead us to
characterize the beam as circularly symmetric, with an effective FWHM of
$23^{\prime}\pm1^{\prime}$.  We use the  pointing information reconstructed from a 
pointing model, which is included in the raw data files, to project 
our simulations onto the sky in the same manner that the real data is
scanned.  

A total of 682 individual hours of experimental data are used 
for the analysis.  The data are naturally divided into 55 minute 
sections by our hourly calibration cycles. These hourly sections are 
a useful size for several reasons.  In addition to the natural delineation 
by calibrations, 55 minutes is a very manageable size for manipulation in 
the \emph{IDL} software package on a desktop computer. Also, sky rotation over 
one hour at our observing angle provides redundant scanning over 
a nearly symmetric sky patch. The most important effect of this choice of time 
slices is on our 'template removal' described below.  
We tested the sensitivity of our results to varying the timescale of our 
template removal from the fiducial hour down to a minimum 
(set by sky rotation) of 600 seconds, and observed no significant changes.  

\begin{figure}  
\plotone{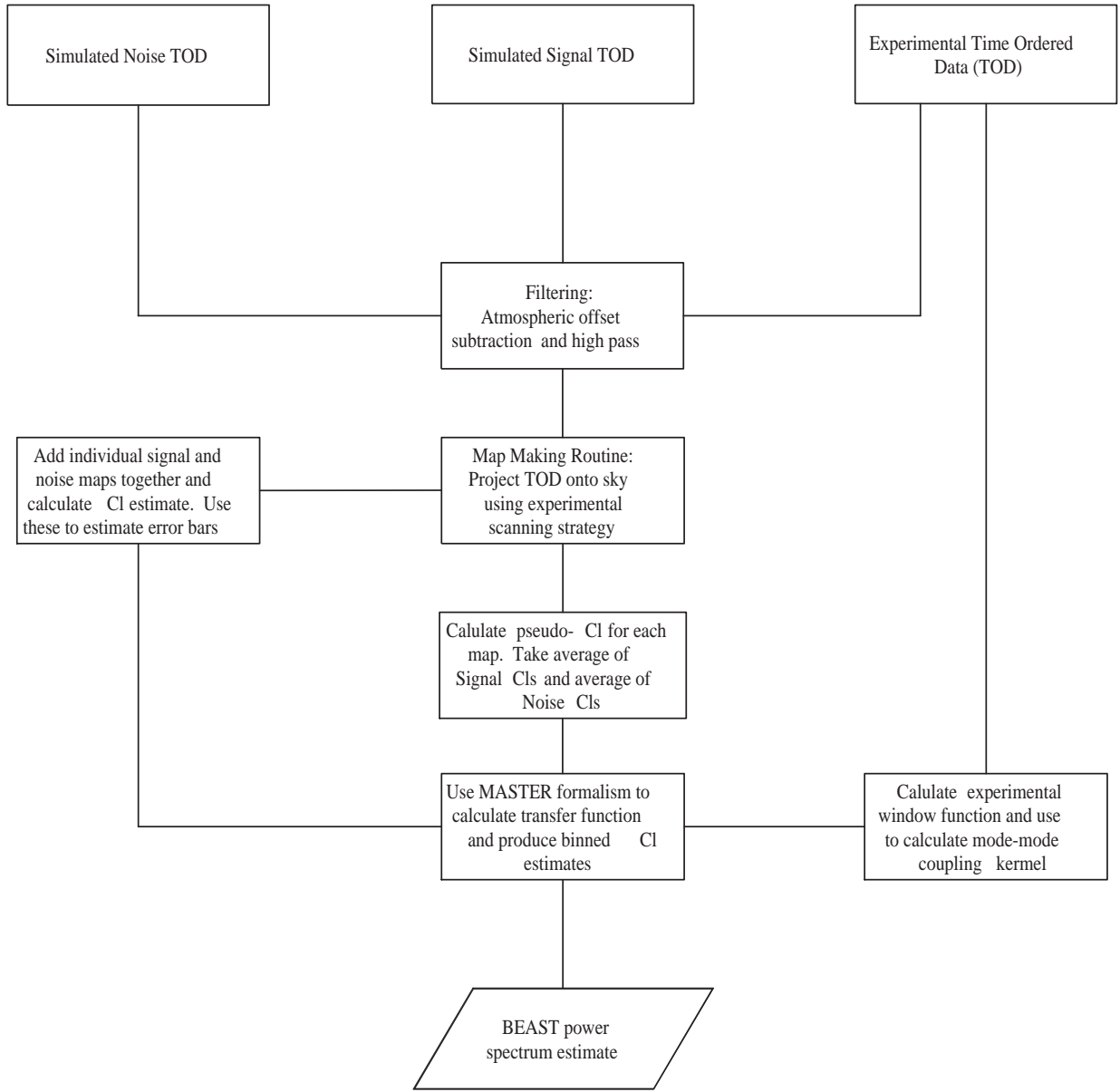}
\caption{\small \sl Overview of the steps in the BEAST simulation and analysis pipeline. }  

\label{fig:flow}  
\end{figure}

The data has been inspected and spurious signal events, e.g. due to aircraft, 
have been removed.  The data includes both the signal measured by the 
experiment and the experimental pointing at that instant.  This information 
is used to construct a sky map of the observed signal.  For all the maps 
created in the data analysis we use the HEALPix 
\footnote[1]{http://www.eso.org/science/healpix/} \citep{healpix} pixelization scheme with an
\emph{nside} parameter of 512.  This results in a map containing 3,145,728
pixels.  Given the size of the experimental beam and the high sampling frequency which is possible with a ground based instrument (450Hz for BEAST), the effects of pixel smoothing are negligible and are ignored here.  For the experimental data we create a HEALPix map and
calculate the CMB power spectrum using the HEALPix \emph{anafast} package.  
Further details of the map making process
can be found in \citet{map_paper}.  Fig. \ref{fig:flow} shows an overview of the steps in the BEAST simulation and analysis pipeline.

The WMAP \citep{wmap} best-fit theoretical power spectrum is used to create 
random realizations of the pure CMB sky.  
We tested the BEAST pipeline with the power spectra from two fiducial 
cosmological models and found the final power spectrum to be unchanged by 
this choice.  The first model was a set of 
reasonable current estimates for cosmological parameters prior to the WMAP 
data release and the second was the best-fit power spectrum published by the 
WMAP team. 

We scan these signal maps using our experimental pointing strategy read 
from the time-ordered data (TOD) files.  We expect the 
time-averaged atmospheric contributions 
to the data to vary with elevation.  To remove this 
foreground we fit a function of elevation angle to the TOD for each hour and subtract it from the TOD samples.  Subsequently a
10Hz high pass filter is used.  The simulation has now been subjected to 
exactly the same scanning and filtering as the real BEAST data and we project 
this simulated data back onto a sky map. 

A foreground mask is applied to remove the Galaxy and point
source contamination from known sources.  We remove from the analysis all  
pixels with latitude $b\le17.5^\circ$.  We tested the analysis pipeline with a range of 
galactic latitude cuts and found that below $b=17.5^\circ$ there was significant 
galactic foreground contamination.  In addition to this, a separate analysis of the Galactic foregrounds for the BEAST experiment \citep{foreground} showed that $b \le 17.5^\circ$ gives an optimal compromise between maximizing the sky fraction observed by the experiment and minimizing the amount of foreground contamination.  In this work it was also found that 
residual Galactic foregrounds outside the mask are small and they are ignored here.  

Finally a power spectrum is generated from each signal map and 
these power spectra are averaged to produce an average signal-only 
power spectrum.

\begin{figure}  
\plotone{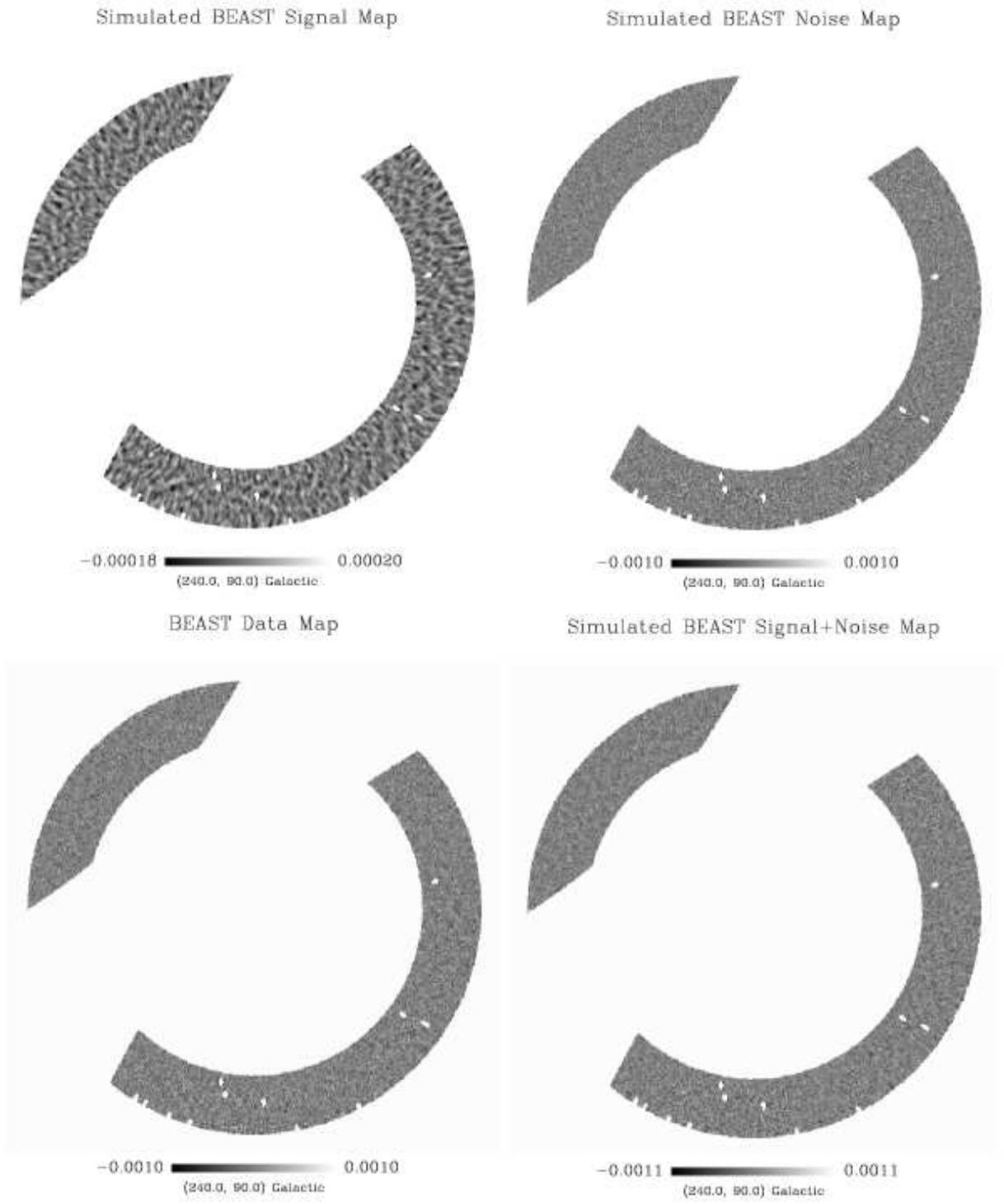}
\caption{\small \sl Comparison of simulated and actual BEAST maps in units of Kelvin.  The noise dominated nature of the BEAST data can be seen by comparing the noise map to the BEAST data map.  }
\label{fig:4_maps}  
\end{figure}

To construct noise-only maps we subtract our signal estimate for the
map from each sample in the experimental TOD and assume that each 
hourly segment of experimental data is now noise-dominated. 
We further assume the noise to be piecewise stationary 
over one hour sections of data and that each one hour noise chunk 
is independent.  We estimate noise power spectra using a windowed FFT
on each hourly segment \citep{NR}. We are then able to generate synthetic 
noise simulations which have the same power spectrum as the actual 
noise from the experiment.  We filter the simulated noise TOD in the 
same manner as for the data and signal simulations and project the noise 
onto a sky map, then calculate the average noise power spectrum.  
Comparisons of the data map and the maps created in the simulation
 pipeline are shown in Fig.\ref{fig:4_maps}.

\begin{figure}  
\plotone{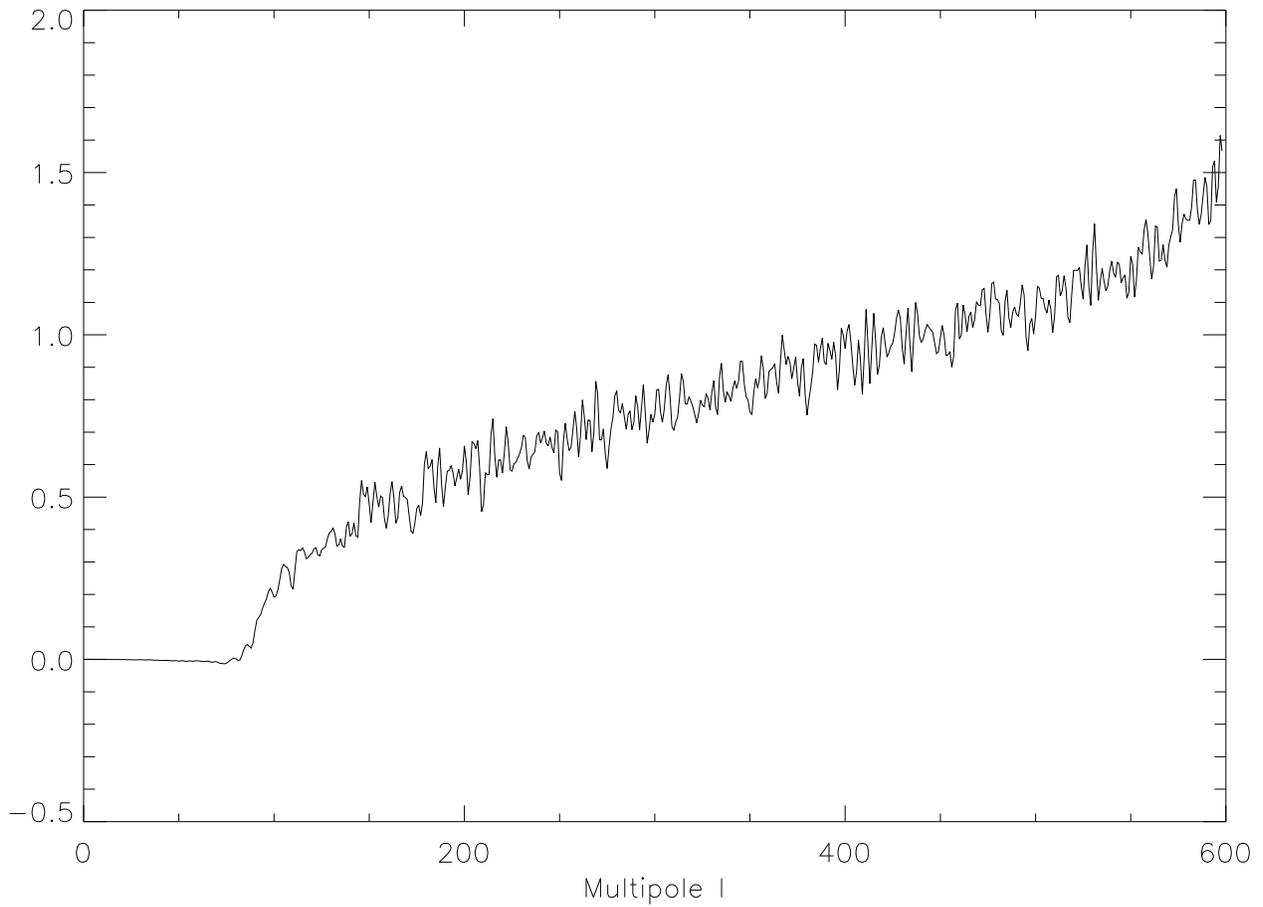}  
\caption{\small \sl Unbinned transfer function for BEAST.  Monte Carlo noise is visible, which is smoothed by the binning process.  The turnover at $\ell \sim$ 550 is caused by the ill conditioned mode-mode coupling kernel.}  
\label{fig:tf}  
\end{figure}

Since we have all of the pointing information, we can also create the
experimental window function on the sky.  This is a simple geometrical
construction which is 1 for any HEALPix pixel which the experiment
observes and 0 elsewhere.  We use this window function to calculate
the mode-mode coupling kernel, $M_{\ell\ell '}$, which depends only on the
geometry of the observed region of sky.  We use the ansatz for the expected
pseudo-$C_\ell$ which was proposed in \citet{Hivon1}.  From the
signal-only simulations we can calibrate the transfer function 

$$F_\ell=M_{\ell\ell'}^{-1}\langle{C_\ell}_s\rangle \langle{C_\ell}\rangle^{-1} (B_\ell^2)^{-1},$$
where $\langle{C_\ell}_s\rangle$ are the signal-only pseudo $C_\ell$ and 
$ \langle{C_\ell}\rangle$ are the best-fit theory $C_\ell$ from the WMAP
experiment.  $B_\ell$ is the experimental beam, a Gaussian with FWHM
of 23 arcmin in this case.  Since the coupling kernel is ill conditioned we
use an iterative approach for computing
$M_{\ell\ell'}^{-1}\langle{C_\ell}_s\rangle$.  The transfer
function for the BEAST experiment is shown  in Fig. \ref{fig:tf}.  

Now our   $C_\ell$ estimate is given by 
$$\hat{C_\ell} = {{M_{\ell\ell'}^{-1} \tilde{C_\ell} - \langle\tilde{N_\ell}\rangle} \over {F_\ell B_\ell^2}}$$
where $\langle\tilde{N_\ell}\rangle$ are the pseudo-$C_\ell$ from the noise Monte Carlo simulations and $\tilde{C_\ell}$ are the pseudo-$c_\ell$ from the data.

\begin{figure}  
\plotone{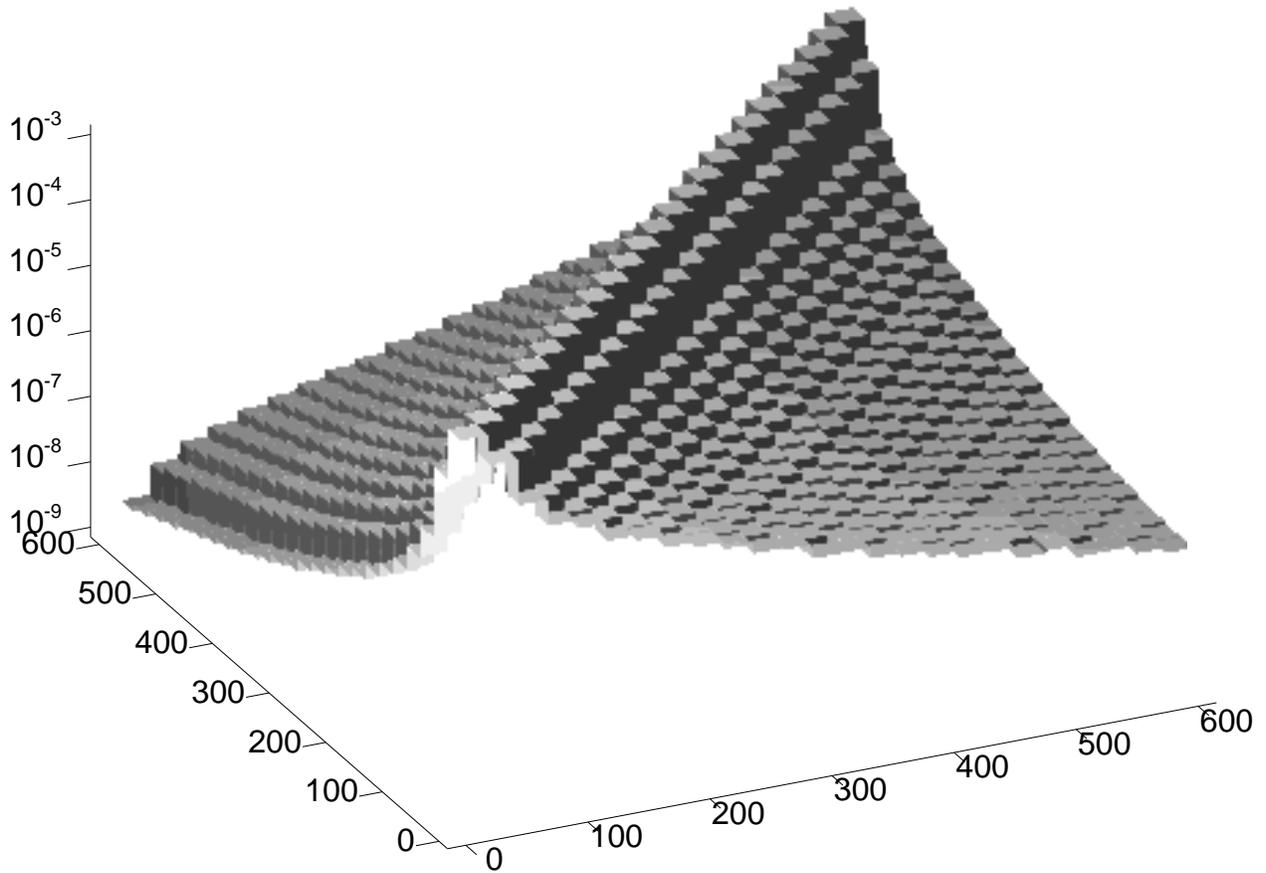}  
\caption{\small \sl Mode-mode coupling kernel for the BEAST experiment.  The z-axis is logarithmically scaled in order to show the off diagonal elements, which decrease rapidly.  The width of the diagonal is approximately 25 in $\ell$ either side of the peak.  In order to avoid correlations between the bins in our final power spectrum, we therefore choose a bin width of 55 in $\ell$. }  
\label{fig:mm}  
\end{figure}

In practice we use the binned
version of the above equation as given in \citet{Hivon1}.  The binned
mode-mode coupling kernel is shown in Fig. \ref{fig:mm}.

By averaging the power spectrum over bins in $\ell$ we effectively reduce correlations 
between the $C_\ell$ bins which were introduced by the sky cut and 
we also reduce the errors on the resulting power spectrum estimator.  We have 
tried different
binning schemes and choose a bin width of  $\Delta\ell=55$.  

Finally, we create sky simulations by adding the signal and the 
noise maps, produced as described above.  The covariance
matrix $C_{bb'}$ of the binned power spectrum is calculated from 
these simulations and the diagonal elements give us the error bars 
on the binned power spectrum estimator. The power spectrum obtained from 
this process is discussed in the next section.

The code for the BEAST analysis pipeline was written and executed 
on an IBM SP RS/6000 (\emph{Seaborg}) at the National Energy Research 
Scientific Computing Center.  The code was
parallelized using MPI and ran on 640 processors.  In order to
obtain a stable PS estimate and to estimate our error bars to $\sim$20\% 
accuracy we required 40 Monte Carlo runs.  The operation count for our
analysis pipeline scales approximately as $N_{tod}log(N_{tod})$ with a
large prefactor, where $N_{tod}$ is the number of samples in the TOD.  

In order to minimize the computational time, we modified the Healpix 
routines \emph{synfast} (which makes a sky map from a power
spectrum) and \emph{anafast} (which calculates the power spectrum from a
sky map) so that they only performed analysis on the portion of the
sky where BEAST scans.  Since the data set read in for the BEAST simulations 
is $\sim$ 80GB and the output maps for 40 MC runs are $\sim$1.7TB, 
we also implemented compression algorithms for storing the output maps on disk.

\section{Power Spectrum and Parameter Estimation}

\begin{figure}[t]  
\plotone{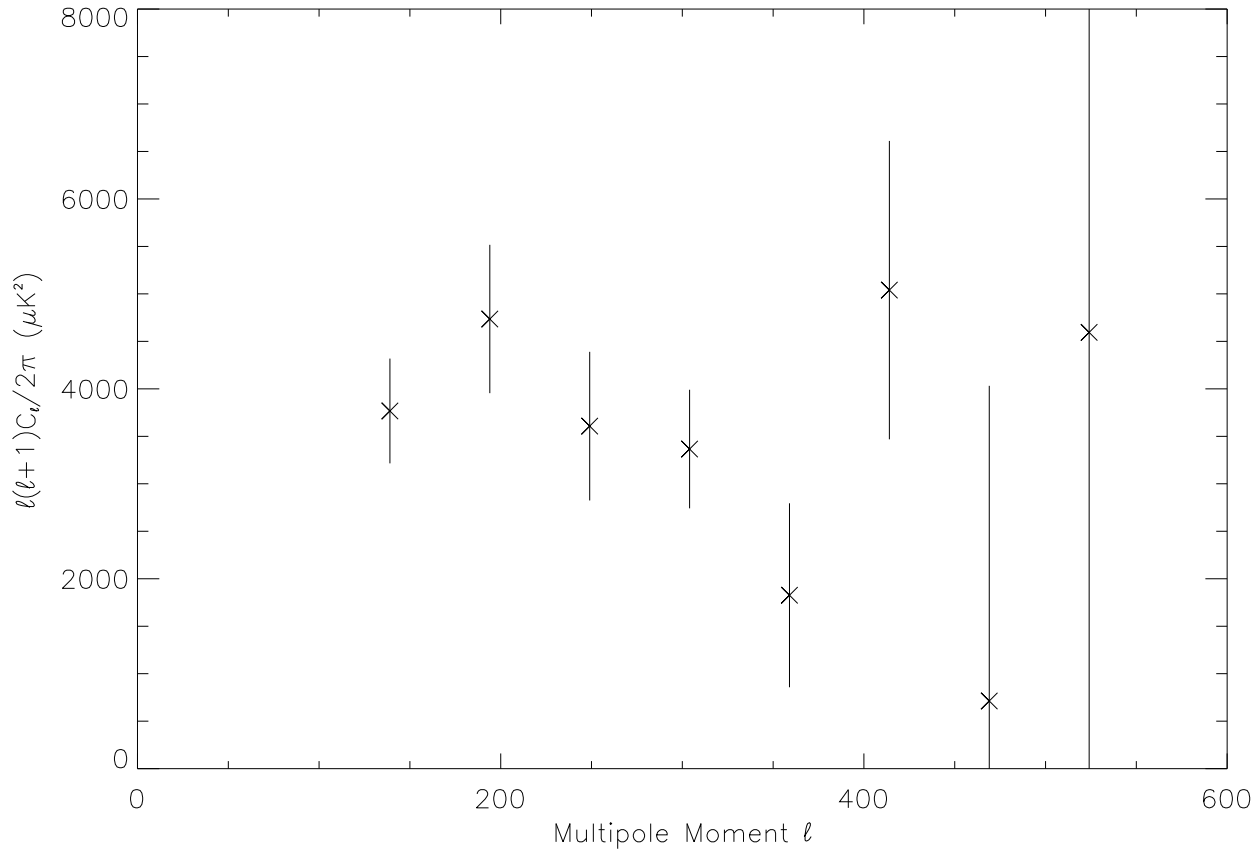}  
\caption{\small \sl CMB anisotropy power spectrum for the BEAST
  experiment.  Error bars are 1$\sigma$}  
\label{fig:pow_spec}  
\end{figure}

The CMB power spectrum extracted from the BEAST data is shown in
 Fig. \ref{fig:pow_spec}.  The values of the power spectrum are shown in
 Table 1 .  The 1-$\sigma$ error bars shown in the figure should be
interpreted
with some caution.  40 Monte Carlo simulations allow us to calculate these
error
bars to within 20\%, which is sufficient for our purposes here, but more
simulations
would lead to more accurate error bars.  In addition, we use the Monte Carlo
simulations to calculate the transfer function ($F_{\ell}$), which is then
used to produce
the $C_{\ell}$ estimates and we use these same simulations to calculate the
error bar on these estimates.  Therefore, our estimate of the error bars on
the power spectrum is not unbiased and we underestimate the size of these
error bars.
In calculating our 8 binned $C_{\ell}$ estimates, we effectively compute a binned transfer function $T_b$ and a binned noise estimate $N_b$ for each bin.  We use 40 Monte Carlo simulations of noise to estimate $N_b$ and 40 signal simulations to estimate $T_b$.  Based on the number of degrees of freedom used to produce these 8 binned $N_b$ and $T_b$, we estimate the bias in the error bar to be approximately 15\%,  
of the same order as our Monte Carlo
uncertainty in the errors. However, since this latter effect is a systematic
bias, the comparison of the BEAST power spectrum estimates and the resulting
parameter estimates to WMAP should be taken as "worst-case" consistency
checks.

\begin{figure}[t]  
\plotone{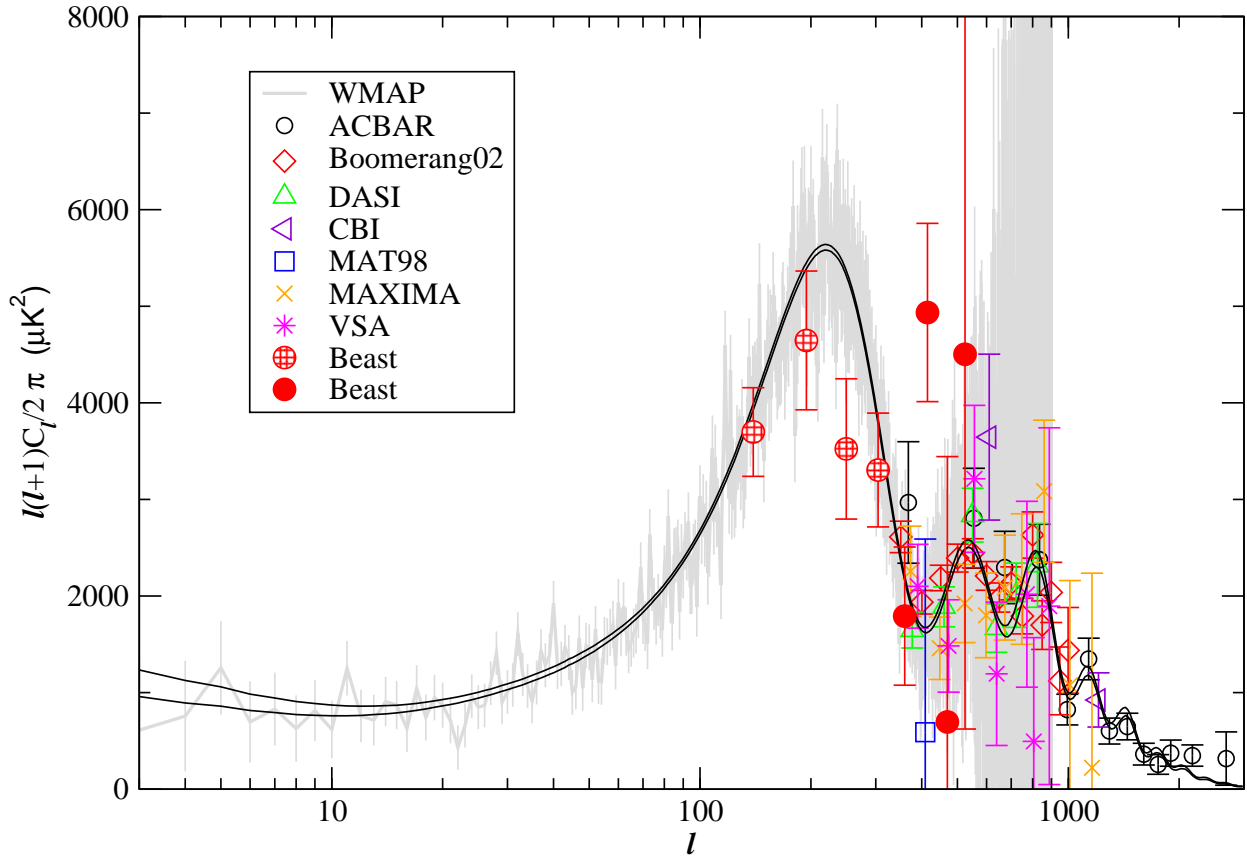}  
\caption{\small \sl Comparison of the BEAST determination of the CMB power spectrum with 8 other recent experiments.  The 4 BEAST points over the first peak are hashed circles.  These points were not used in the parameter estimation since they overlap with WMAP, which is cosmic variance limited over this range.  The remaining 4 BEAST data points are solid circles.}  
\label{fig:ps_comp}  
\end{figure}

A $\chi^2$ comparison of the BEAST data and the WMAP data was
performed.  For this comparison the WMAP data was assumed to
have zero error.  We find a $\chi^2$ parameter of 15.02.  With 9 degrees of
freedom this means a larger value of $\chi^2$ would occur approximately 10\%
of the time, so the BEAST power spectrum is marginally consistent with the
WMAP
result.  We show the BEAST power spectrum overplotted with the power spectra
from several recent experiments in Fig. \ref{fig:ps_comp}.

After the mean power spectrum was determined, its likelihood was sampled
40 times, producing 40 sample binned power spectra. The likelihood around
the power spectrum is not, in general, Gaussian distributed, but through
a change of variables - to the log-offset-normal variables of
Bond, Jaffe and Knox (BJK parameterization \citep{BJK}) - the distribution can
be
mapped into one that is much more nearly Gaussian. However, it was found
that 40 samples of the power spectrum distribution was too few for a
reliable determination of the BJK parameters, and thus it was decided
that the power spectrum likelihood would be approximated as
Gaussian-distributed. We then calculate the Likelihood $L$ of a theoretical
power
spectrum, $D^{th}_i$, as follows:

 $  \chi^2 = \sum_{ij} (D^{th}_i - D^{ob}_i) M_{ij} (D^{th}_j - D^{ob}_j) $

 $  L = \exp(- \chi^2 / 2) $

 $    D^{ob}_i \equiv C^{ob}_i l(l+1)/2\pi $

where $C^{ob}_i$ is the observed band-power of the i-th bin, and $M_{ij}$ is
 the covariance matrix.

 We determined the best-fit (maximally likely) points in parameter space for:

 1)  BEAST data + WMAP + MAXIMA1,MAT98,BOOMERANG02,DASI01,VSA1,ACBAR1,CBI +
 Hubble Key Project + Big Bang Nucleosynthesis relation between $\Omega_B h^2$ and $He^4$
 mass-fraction ($Y_p$) \citep{hcw} over the parameter space:
     $\Omega_m$, $\Omega_{\Lambda}$, $h$, $n_s$, $\Omega_B h^2$, $Y_p$,
$\tau$,
 $n_t$, $r$

 2)  BEAST data alone.

 For 1) BEAST data + other recent cosmological data, we found the parameter values
and errors via a Markov chain approach. Starting from a 30000 point Markov chain
previously run with the experiments: WMAP + MAXIMA1,MAT98,BOOMERANG02,DASI01,VSA1,ACBAR1,CBI +
 Hubble Key Project + BBN  $\Omega_B h^2$-$He^4$ relation, the Markov chain was
thinned by discarding 99 out of every 100 points. Each point was then weighted
by the Beast likelihood. From this weighted point distribution the parameter means
and variance matrix were determined. The parameter estimates were taken to be the means,
and the parameter errors were taken as square roots of the diagonal elements of
the variance matrix.

For 2) BEAST data alone,
 cosmic parameter space was searched for the maximally likely point by first
 trying several candidate points, and then applying the Numerical Recipes
 Amoeba algorithm \citep{NR} to minimize the trial $\chi^2$.  The Amoeba
 algorithm has no inherent minimum scale (similar to adaptive mesh refinement,
  the resolution increases as necessary, with the precision limited only by
 the machine floating point arithmetic), and makes no assumptions about the
 shape of likelihood function.

 Once the BEAST-alone best-fit cosmic parameters have been found, we determined the errors
  in these values. Ideally a method that, again, does not depend on the
 parameter likelihood function having a particular shape (ie: Gaussian,
 for example), such as a Markov Chain algorithm, would be used. In this case
 however, a less computationally costly method can be employed.  We determined
 the errors in the best-fit parameter values by fitting the likelihood
 function around that point to a multivariate Gaussian. The resulting estimate
 of the errors is crude, but sufficient to give an overall measure of the
dispersion.
To the extent to which the likelihoods are approximately Gaussian in the
narrowly constrained case (1) we expect these errors to be more accurate.
The results of the
 joint parameter estimation for BEAST plus other experiments are shown in
 Table 2.  We found the BEAST alone parameters to be consistent with these
 values, although much less well constrained.  For example we found
 $\Omega_k\equiv 1-\Omega_{tot}=-0.074 \pm 0.070$ for BEAST alone compared
 with $-0.014 \pm 0.012$ for the joint estimate.

\begin{figure}[t]  
\plotone{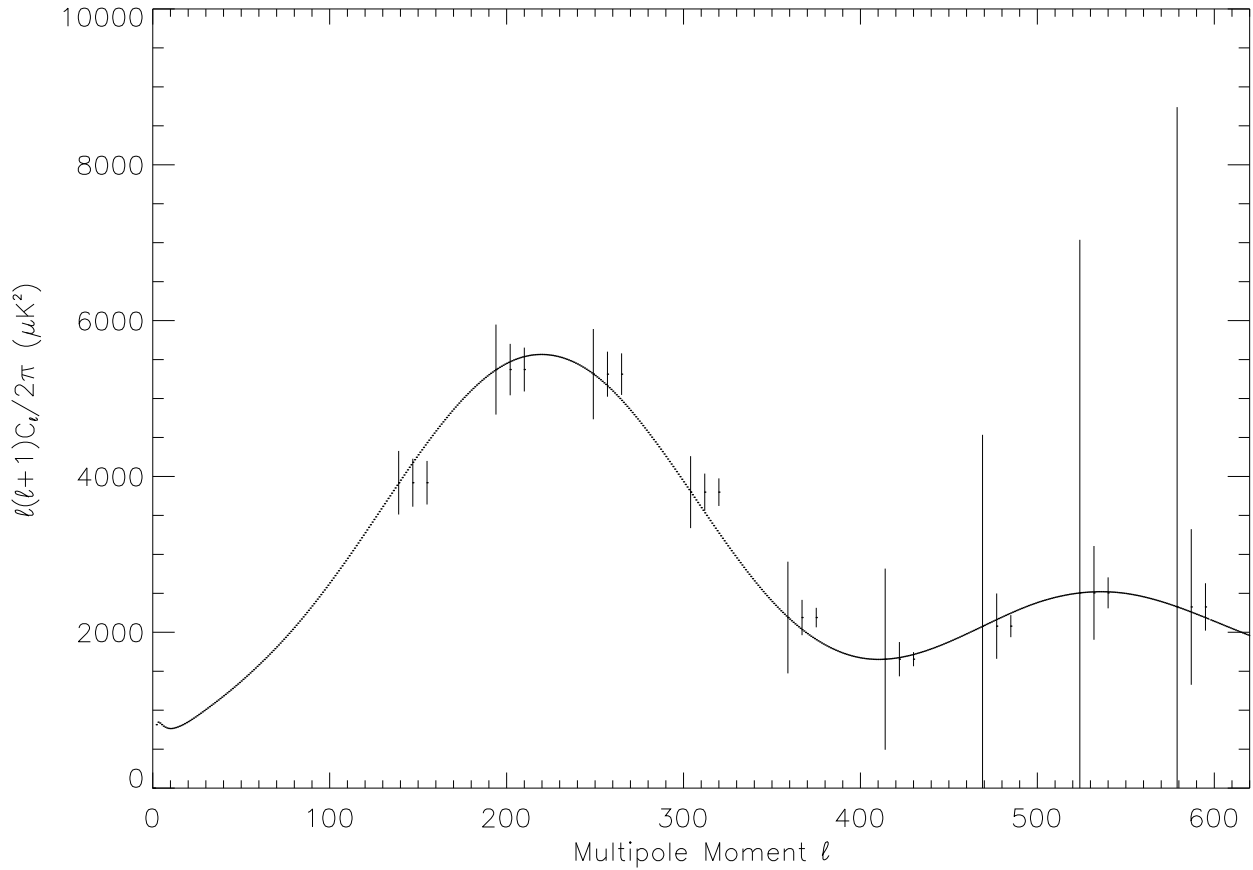}  
\caption{\small \sl We examined the effect on the power spectrum error
  bars of increasing the quantity of data.  4 and 16 times more data
  were considered, effectively reducing the noise by factors of 2 and
  4 respectively.  The original error bars are plotted, followed by
  the half and quarter noise error bars.  The original error bars are
  centered on the $\ell$ bin, while the half and quarter noise are offset
  from the original position for illustrative purposes. In
  the analysis all of the error bars were calculated at the same $\ell$.}  
\label{fig:error_bars}  
\end{figure}

In order to examine possible future directions for the BEAST
experiment we analyzed the effect of increased quantities of data on
the power spectrum error bars.  A two- and four-fold reduction in the
simulated noise were considered, equating to four and sixteen times
more data respectively, assuming no improvement in radiometer sensitivity.  We found that over the first peak in the
power spectrum there was not a significant improvement in the error
bars with more data (see Fig \ref{fig:error_bars}).  This is expected, since in this region we are
sample variance limited by the relatively small patch of sky we observe.  However, at
larger $\ell$ we do see a significant improvement in the power spectrum error
bars, up to the point where the experimental beam cuts off around an
$\ell$ of 600, when the error bars become large regardless of the
amount of data.

\section{Conclusions}

We have presented the angular power spectrum of the cosmic microwave
background as measured by the BEAST experiment.  We have demonstrated 
that it is possible to extract cosmological signal from
an easily accessible, ground based CMB experiment which is dominated 
by correlated noise and that the resulting power spectrum and parameter 
estimation is consistent with previous results.  

The MASTER method was successfully implemented and although this method 
is approximate, it proved to be flexible and robust and 
produced a power spectrum with less than 1000 CPU hours of computational 
time.  We believe the BEAST CMB dataset to be one of the largest 
TOD's analyzed to date and 
this proved feasible within the MASTER framework.  This suggests that the 
analysis of future, larger CMB datasets (e.g. Planck) should be 
computationally feasible.

We also analyzed how additional observing time would improve the power spectrum errors 
and found that significant improvements could be made 
between $250 \le l \le 600$ with additional time.  We note that the atmospheric conditions 
at White Mountain allow for a better than 50\% 'good observing' fraction over the year and that the 26 days of data presented here were limited by funding constraints.

\acknowledgements{}

This work was partially supported by the University of Illinois at 
Champaign-Urbana.  This work has been partially supported by the National 
Computational Science Alliance under grant number AST020003N.  This work was funded by NASA grants NAG5-4078, NAG5-9073 ,
and NAG5-4185 , and by NSF grants
9813920 and 0118297. In addition we were supported by the
White Mountain Research Station, the California Space
Institute (CalSpace), and the UCSB Office of Research.  This research 
used resources of the National Energy Research Scientific Computing Center, which 
is supported by the Office of Science of the U.S. Department of Energy under 
Contract No. DE-AC03-76SF00098.   
The research described in this paper was carried out in part at the Jet 
Propulsion Laboratory, California Institute of Technology, under a contract 
with the National Aeronautics and Space Administration.  J.M is
supported by FAPESP grants 01/13235-9 and 02/08471-1. T.V. and
C.A.W were partially supported by FAPESP grant 00/06770-2. T.V was
partially supported by CNPq grants 466184/00-0 and 302266/88-7-FA. CAW
was partially supported by CNPq grant 300409/97-4-FA and FAPESP grant
96/06501-4.  We acknowledge the use of the Legacy Archive for Microwave Background 
Data Analysis (LAMBDA). Support for LAMBDA is provided by the NASA 
Office of Space Science.  N.F and A.P. were partially supported by CNPq grant number 470531/2001-0.  B.D.W. acknowledges the 2003/4 NCSA Faculty Fellowship. 
Some of the results in this paper have been derived using HEALPix \citep{healpix}.  
We would like to thank Julian Borrill at NERSC for valuable discussions on the
computational aspects of this project.  We also thank members of the Planck community 
for stimulating discussions.  

\begin{deluxetable}{llcc}
\tablecaption{Beast Power Spectrum Estimates}
\tablewidth{0pt}
\scriptsize
\tablehead{
  Bin       &  Bin     &           Estimate in  $\mu K^2$  of & 1-$\sigma$    \\
$\ell_{min}$  &  $\ell_{max}$  &  $\ell(\ell+1)C_{\ell}/2\pi$ & error     \\
}
\startdata
139	&     193   &   3776  &   $\pm$552                 \\
194	&     248   &   4744  &   $\pm$781                  \\
249     &     303   &   3597  &   $\pm$782                  \\
304     &     358   &   3374  &   $\pm$625                  \\
359     &     413   &   1829  &   $\pm$969                  \\
414     &     468   &   5040  &   $\pm$1571                  \\
469     &     523   &   711   &   $\pm$3319                 \\
524     &     678   &   4599  &   $\pm$6136                  \\
\enddata
\tablecomments{\small \sl The BEAST $C_{\ell}$ estimates obtained using the MASTER method.  The starting and ending values of each $\ell$ bin are shown.  The $C_{\ell}$ values in the table and those shown in Fig. \ref{fig:pow_spec} are averaged over these bins.}
\end{deluxetable}

\begin{deluxetable}{lc}
\tablecaption{Cosmological Parameter Estimates}
\tablewidth{0pt}
\scriptsize
\tablehead{
Parameter               &  BEAST+others        \\
}
\startdata
$\Omega_k$		&   -0.014$\pm$0.011       \\
$\Omega_{CDM}h^2 $      &   0.094$\pm$0.012     \\
$\Omega_bh^2$           &   0.024$\pm$0.002        \\
$h$                     &   0.727$\pm$0.048        \\
$n_s$     	        &   1.002$\pm$0.052        \\
$\tau$                  &   0.154$\pm$0.074    \\
$Y_p$                   &   0.249$\pm$0.001     \\
\enddata
\tablecomments{\small \sl BEAST parameter estimates calculated using a joint
analysis with other CMB data and BBN and Hubble Key Project constraints.
$\Omega_k\equiv 1-\Omega_{tot}$.  The parameter errors were obtained from the variance of a Markov chain in parameter space.}
\end{deluxetable}


\begin{thebibliography}{}

\bibitem[Bennett et al. (2003)]{wmap} 
Bennett, C.~L.~et al. 2003, \apjs, 148, 1

\bibitem[Childers et al. (2003)]{childers}
Childers, J. et al., 2003, in preparation

\bibitem[Figueiredo et al. (2003)]{optics}
Figueiredo, N. et al, 2003, in preparation

\bibitem[G{\' o}rski, Hivon \& Wandelt (1999)]{healpix}
G{\' o}rski, K.~M., Hivon, E., \& Wandelt, B.~D.\ 1999, Analysis Issues for Large CMB Data Sets.  Proceedings: Evolution of Large Scale Structure, Garching. 


\bibitem[Grainge et al. (2002)]{vsa}
Keith Grainge \emph{et al.} Mon. Not. R. Astron. Soc. 000, 1­5 (2002)

\bibitem[Halverson et al. (2001)]{dasi}
Halverson, N. W. \emph{et al.}, 2001, astro-ph/0104489

\bibitem[Hanany et al. (2000)]{maxima}
Hanany, S. \emph{et al.}\ 2000, \apjl, 545, L5

\bibitem[Hivon et al. (2002)]{Hivon1}Hivon, E., G{\' o}rski, K.~M., Netterfield, C.~B., Crill, B.~P.,
Prunet, S., \& Hansen, F.\ 2002, \apj, 567, 2

\bibitem[Huey, Cyburt \& Wandelt (2003)]{hcw}Huey, G, Cyburt, R. H., Wandelt, B.D., 2003, astro-ph/0307080, Phys. Rev. D, in press

\bibitem[Knox et al. (1998)]{BJK} Knox, L., Bond, J.~R., 
Jaffe, A.~H., Segal, M., \& Charbonneau, D.\ 1998, \prd, 58, 83004 

\bibitem[e.g. Kolb and Turner (1994)]{KT94}
Kolb, E.W. and Turner, M.S., The Early Universe, 1994, Addison-Wesley

\bibitem[Kuo et al.(2002)]{acbar} Kuo, C.~L.~et al.\ 2002, 
Bulletin of the American Astronomical Society, 34, 649 

\bibitem[Mej\'{\i}a et al. (2003)]{foreground}
Mej\'{\i}a, J. et al., 2003, In Preparation

\bibitem[Meinhold et al. (2003)]{map_paper}
Meinhold, P. M, et al, ApJ, submitted.  (astro-ph/0302034)

\bibitem[Miller et al. (1999)]{toco98}Miller, A.D. \emph{et al.} 1999 Astrophys.J. 524 (1999) L1-L4

\bibitem[Padin et al. (2001)]{cbi} Padin, S. \emph{et. al.}, ApJ 549, L1, (2001)

\bibitem[e.g. Peebles (1993)]{peebles} Peebles, P.J.E., Principles of Physical Cosmology, 1993, Princeton University Press

\bibitem[Press et al. (1986)]{NR}
Press, W.H., Flannery, B.P.,  Teukolsky, S.A., \& Vetter, W.T., Numerical Recipes - The Art of Scientific Computing, Cambridge University Press, Cambridge 1986

\bibitem[Ruhl et al. (2002)]{boomerang02}Ruhl, J.E. \emph{et al.} 2002, astro-ph/0212229

\bibitem[Wandelt, Hivon \& G\'orski (2001)]{whg} Wandelt, B.D., Hivon, E.F, G\'orski, K.M. 2001, Phys. Rev. D64, 083003

\end{thebibliography}
\end{document}